\documentclass[11pt]{article}
\usepackage{graphicx}
\usepackage{epstopdf}

\begin{document}

\def\xslash#1{{\rlap{$#1$}/}}
\def \p {\partial}
\def \dd {\psi_{u\bar dg}}
\def \ddp {\psi_{u\bar dgg}}
\def \pq {\psi_{u\bar d\bar uu}}
\def \jpsi {J/\psi}
\def \psip {\psi^\prime}
\def \to {\rightarrow}
\def\bfsig{\mbox{\boldmath$\sigma$}}
\def\DT{\mbox{\boldmath$\Delta_T $}}
\def\xit{\mbox{\boldmath$\xi_\perp $}}
\def \jpsi {J/\psi}
\def\bfej{\mbox{\boldmath$\varepsilon$}}
\def \t {\tilde}
\def\epn {\varepsilon}
\def \up {\uparrow}
\def \dn {\downarrow}
\def \da {\dagger}
\def \pn3 {\phi_{u\bar d g}}

\def \p4n {\phi_{u\bar d gg}}

\def \bx {\bar x}
\def \by {\bar y}

\begin{center} 
{\Large\bf   Matching of Fracture Functions for SIDIS in Target Fragmentation Region   }
\par\vskip20pt
K.B. Chen $^{1,2}$, J.P. Ma$^{2,3,4}$ and X.B. Tong$^{5,6,2}$    \\
{\small {\it
$^1$ School of Science, Shandong Jianzhu University, Jinan, Shandong 250101, China\\
$^2$ CAS Key Laboratory of Theoretical Physics, Institute of Theoretical Physics, P.O. Box 2735, Chinese Academy of Sciences, Beijing 100190, China\\
$^3$ School of Physical Sciences, University of Chinese Academy of Sciences, Beijing 100049, China\\
$^4$ School of Physics and Center for High-Energy Physics, Peking University, Beijing 100871, China\\
$^5$ School of Science and Engineering, The Chinese University of Hong Kong, Shenzhen, Shenzhen, Guangdong, 518172, China\\
$^6$ University of Science and Technology of China, Hefei, Anhui, 230026, China }} \\
\end{center}
\vskip 1cm
\begin{abstract}
In the target fragmentation region of Semi-Inclusive Deep Inelastic Scattering, the diffractively produced hadron has small transverse momentum. If it is at order of $\Lambda_{QCD}$,   it prevents to make predictions 
with the standard collinear factorization. However, in this case, differential cross-sections can be predicted by the factorization with fracture functions, diffractive parton distributions.  If the transverse momentum is much larger than  $\Lambda_{QCD}$ but much smaller than $Q$ which is the virtuality of the virtual photon, both factorizations apply. In this case, fracture functions  can be factorized with collinear parton distributions and fragmentation functions. We study the factorization up to twist-3 level and obtain gauge invariant results. 
 They will be helpful for modeling fracture functions and useful for resummation of large logarithm of the transverse momentum appearing in collinear factorization. 
   
\vskip 5mm
\noindent
% PACS numbers
\end{abstract}
\vskip 1cm

\noindent
{\bf 1. Introduction} 
\par\vskip5pt

Semi-Inclusive DIS(SIDIS) have three different kinematical regions in which different QCD factorizations are used. These regions are classified according to the momentum of the observed hadron in the final state. 
In the region where the transverse momentum is large, the production rate can be predicted with standard QCD collinear factorization. In the region where the observed hadron moves in the forward region of the virtual photon with a small transverse momentum, one can use Transverse-Momentum-Dependent QCD factorization\cite{CSS,JMY}, which involves TMD parton distributions and TMD parton fragmentation functions. 
The third region is called as target fragmentation region, where the observed hadron moves in the forward region of the initial hadron with a small transverse momentum $k_\perp$. In this region, the standard collinear  
factorization fails because of that the perturbative part in the factorization becomes divergent as powers of $\ln k_\perp$. 

\par 
Experimentally, the production in the target fragmentation region has been observed in HERA experiment\cite{HERA}. It has been suggested in \cite{TrVe} that the production rate can be factorized 
with fracture functions.  An one-loop study of SIDIS in \cite{DGSIDIS} shows that the factorization holds at the order.
One can prove  the factorization with fracture functions at all orders, as discussed in \cite{BeSo, DIFCOL}.  For the production of one hadron combined with
a lepton-pair in hadron collisions,  factorizations with fracture functions  have been shown to hold at one-loop level in \cite{FCLT, FC}, where the produced hadron 
is in the forward- or backward regions. However, the factorization for hadron collisions can be failed  as discussed in \cite{CFS}. 
Fracture functions, also called as diffractive parton distributions,  are parton distributions of an initial hadron with one diffractively produced  particle observed in the final state.  Currently, most information about fracture functions comes from analysis of HERA data, e.g., in \cite{HERADF1, HERADF2}. The production in target fragmentation region 
can be studied with current experiments at  JLab  and planned experiments at Eic\cite{Eic} and EicC\cite{EicC}. 
Hence, more information of fracture functions will be available.

It is noted that the factorization in the target fragmentation region holds for $Q\gg k_\perp$ where $Q$ is the square root of the momentum transfer of the virtual photon of SIDIS. In the case of $Q \gg k_\perp\gg \Lambda_{QCD}$, one can also use the collinear factorization to factorize the differential cross-sections
with collinear parton distributions and fragmentation functions. Therefore, for $k_\perp\gg \Lambda_{QCD}$,
fracture functions in SIDIS can be factorized with collinear parton distributions and fragmentation functions.    
In this paper, we study the factorization up to twist-3. The matching of twist-2 fracture functions is straightforward, while the matching of twist-3 fracture functions is nontrivial in the sense that one has to keep the gauge invariance. Here we will mainly study the factorization or matching at the level of twist-3.  

Fracture functions at twist-2 relevant to SIDIS have been defined in \cite{BeSo}, where their asymptotic behavior 
has been derived for the region where the momentum fraction of the struck parton approaches 
its maximal value.  In \cite{ABAK} TMD quark fracture functions have been classified for a polarized 
spin $1/2$ hadron in the initial state.  An one-loop matching of various twist-2 fracture functions for the process where the production of a lepton pair with large invariant mass in hadron collisions associated with a diffractively produced photon instead of a hadron in different kinematical regions,  has been studied in \cite{FFP}. 

In the target fragmentation region  with $Q\gg k_\perp$  SIDIS can be conveniently described with fracture functions. If one uses collinear factorization in the region of $Q \gg k_\perp\gg \Lambda_{QCD}$, then there will be terms with large log's like $\ln Q/k_\perp$ in perturbative coefficient functions. Using the factorization with fracture functions and the matching studied here, these terms with large log's can be re-summed with helps of evolution equations.  In collinear factorization, twist-3 effects involve in general 
twist-3 parton distributions and twist-3 parton fragmentation functions. 
 It is interesting to notice that the latter is absent in the matching of twist-3  studied here. The reason for this 
 is helicity conservation in QCD as we will show.

The content of our paper is: In Sect.2. we introduce our notations and definitions of relevant parton distributions, fragmentation functions and fracture functions. The results of matching of twist-2 fracture functions are also given in this section. In Sect.3. we derive the matching of the twist-3 fracture function which is relevant to single transverse-spin asymmetry. In Sect.4. the matching of the twist-3 fracture function 
responsible to a double-spin asymmetry is given. Sect.5 is our summary.

\par\vskip20pt
\noindent
{\bf 2. Notations and Matching of Twist-2 Fracture Functions} 

\par\vskip5pt
We will use the  light-cone coordinate system, in which a
vector $a^\mu$ is expressed as $a^\mu = (a^+, a^-, \vec a_\perp) =
((a^0+a^3)/\sqrt{2}, (a^0-a^3)/\sqrt{2}, a^1, a^2)$ and $a_\perp^2
=(a^1)^2+(a^2)^2$. We introduce two light cone vectors $n^\mu = (0,1,0,0)$ and  $l^\mu =(1,0,0,0)$. The transverse metric is given by $g_\perp^{\mu\nu} = g^{\mu\nu} - n^\mu l^\nu - n^\nu l^\mu$. We will also need the transverse  antisymmetric tensor which is given by   
$\epsilon_\perp^{\mu\nu} =  \epsilon^{\alpha\beta \mu\nu } l_\alpha n_\beta$ with $\epsilon_\perp^{12}=1$.
Throughout of the work, we will use Feynman gauge.

\par\vskip15pt
\noindent 
{\bf 2.1. Definitions of Fracture Functions and SIDIS in Target Fragmentation Region} 
\par  
We consider an initial hadron $h_A$ with the momentum $P^\mu = (P^+, P^-,0,0)$ and another hadron in the final state with the momentum $k^\mu =(k^+, k^-, k_\perp^1, k_\perp^2)$ with $k^+= \xi P^+$. 
The transverse momentum $k_\perp$ is small in comparison with $k^+$.  
The quark fracture functions or quark diffractive parton distributions can be defined with the density matrix: 
\begin{eqnarray}
{ \mathcal M}_{F ij}(x, \xi, k_\perp)  =  \int \frac{ d\lambda  } {2\pi} e^{-i x P^+ \lambda }
\sum_X \langle h_A  \vert  \biggr [ \bar q(\lambda n ) {\mathcal L}_n^\dagger (\lambda n) \biggr ]_j  \vert X h  \rangle \langle h X \vert 
 \biggr [ {\mathcal L}_n (0) q (0)\biggr ]_i \vert h_A  \rangle  , 
\label{DENM}
\end{eqnarray}
with $ij$ stand for Dirac- and color indices. Here the transverse momentum of the struck quark is integrated over. To make the density matrix gauge-invariant, the gauge links 
along the $n$-direction are implemented: 
\begin{equation} 
  {\mathcal L}_n (x) = {\rm P} \exp \biggr \{ - i g_s \int_0^\infty d\lambda G^{+}(\lambda n +x)  \biggr \}. 
\end{equation}      
The decomposition of the density matrix has been done in \cite{ABAK}. 
We consider the case that the initial hadron is a spin-1/2 particle and the polarization of the final 
hadron is summed if it has nonzero spin. The spin vector of the initial hadron can be decomposed 
as: 
\begin{equation} 
  s^\mu = s_L \frac{  l^\mu P\cdot n - n^\mu P\cdot l} {m_A} + s_\perp^\mu, \quad\quad P^2 =m_A^2
\end{equation}   
with $s_L$ as the helicity.  Neglecting terms beyond twist-3, the density matrix can be decomposed as: 
\begin{eqnarray} 
  {\mathcal M}_{Fij}(x,  k ) &=& \frac{1}{2 N_c}  \biggr ( \gamma^-  \biggr )_{ij}  \biggr [ 
    F_q (x,\xi, k_{\perp} ) + \epsilon_\perp^{\mu\nu} k_{\perp\mu} s_{\perp \nu} \frac{1}{m_A} 
      F_{qT} (x,\xi, k_{\perp})  \biggr ]    
\nonumber\\ 
  && + \frac{1}{2N_c} \biggr ( \gamma_5 \gamma^-  \biggr )_{ij}  \biggr [ 
    s_L \Delta  F_{q}  (x,\xi, k_{\perp} ) + k_{\perp} \cdot s_{\perp } \frac{1}{m_A}\Delta F_{q T} (x,\xi, k_{\perp} )  \biggr ] +\cdots,
\label{DEFQ}    
\end{eqnarray} 
where $\cdots$ denote those distributions defined with chirality-odd operators. These distributions 
will not contribute to SIDIS in target fragmentation region.  The function $F_q$ and $\Delta F_q$ are of twist-2, 
while $F_{qT}$ and $\Delta F_{qT}$ are of twist-3. We neglect here the contributions beyond twist-3.

\begin{figure}[hbt]
	\begin{center}
		\includegraphics[width=8cm]{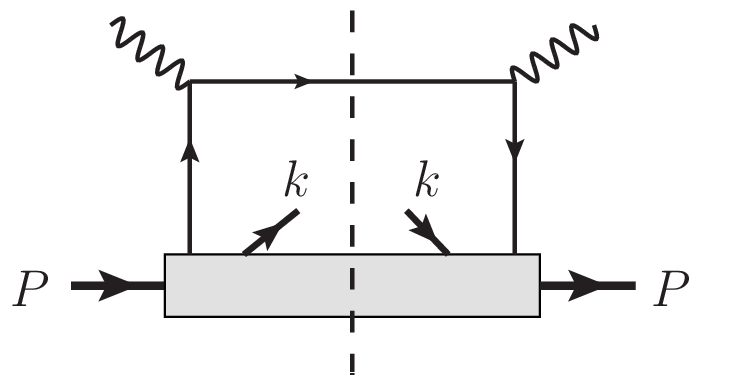}
	\end{center}
	\caption{Diagram for SIDIS in target fragmentation region   }
	\label{SIDISFig1}
\end{figure} 
\par 
 We consider the SIDIS process: 
\begin{equation} 
   e(k_e,\lambda_e) + h_A (P,s) \to e(k_e') + h(k) + X
\label{Eq1} 
\end{equation} 
where the initial hadron is of spin-1/2 with the spin vector $s$. The initial electron can be 
polarized with the helicity $\lambda_e$. We assume that the polarization of the hadron in the final state is not observed. At leading order of QED, there is an exchange of one virtual photon with the momentum $q=k_e-k_e'$ between the electron and the initial hadron. 
The relevant hadronic tensor is:
\begin{equation} 
W^{\mu\nu} = \sum_X \int \frac {d^4 x}{(2\pi)^4} e^{iq\cdot x} \langle h_A \vert J^\mu (x) \vert h X\rangle 
     \langle X h \vert J^\nu (0) \vert h_A \rangle. 
\end{equation} 
The standard variables for SIDIS are:
\begin{equation} 
 Q^2 = -q^2, \quad x_B = \frac{Q^2}{2 P\cdot q},\quad  y=\frac{ P\cdot q}{P\cdot k_e}, \quad  z_h=\frac{P\cdot k}{P\cdot q}. 
\end{equation} 
The masses of hadrons and leptons are neglected.  We take a frame in which the momenta of the initial hadron and the virtual photon are:
\begin{equation} 
   P^\mu \approx ( P^+, 0,0,0), \quad q^\mu =(q^+,q^-, 0,0),
\label{C1}     
\end{equation} 
i.e., the hadron moves in the $z$-direction, and the virtual photon moves in the $-z$-direction.
We consider the case in which the transverse momentum $k_\perp$ is small with $k_\perp^2 \ll Q^2$.  
In the region $z_h \sim {\mathcal O}(1)$, the produced hadron moves almost in the forward region of the virtual photon. 
This region is  called as current fragmentation region. Transverse-Momentum-Dependent(TMD)  
factorization applies in this region.  In the region of $z_h \ll 1$, the produced hadron moves almost in the forward region of the initial hadron. This region is called as target fragmentation region. In this region the hadronic tensor 
can be factorized with the above introduced fracture functions.  

\par 
In the target fragmentation region, we write the momentum of the produced hadron as:
\begin{equation} 
   k^\mu = \xi P^+ l^\mu + k^- n^\mu  + k_\perp^\mu, \quad k^- = \frac{k_\perp^2}{2 \xi P^+}, \quad 
     z_h = \frac{ x_B k_\perp^2}{\xi Q^2}.  
\end{equation}    
We consider the experimental situation in which the initial hadron is polarized and the transverse part 
 $s_\perp^\mu$ of  the spin vector $s^\mu$ is nonzero.  The incoming- and outgoing lepton span the so-called lepton plane. In our frame specified in the above  
the azimuthal angle between the spin vector $s_\perp$ and the lepton plane is denoted $\phi_s$. Similarly, one defines the azimuthal angle $\phi_h$ for the produced hadron.
The azimuthal angle of the outgoing lepton around the lepton beam with respect to the spin vector is denoted $\psi$.
In the kinematical region of SIDIS with large $Q^2$, one has $\psi\approx \phi_s$\cite{Diehl}. With this specification the differential cross-section is given by \cite{Diehl,NSIDIS}:
\begin{equation} 
    \frac{ d\sigma}{d x_B d y d \xi  d\psi d^2  k_{\perp} } = \frac{\alpha^2 y}{4 \xi Q^4}   L_{\mu\nu} W^{\mu\nu}, 
\end{equation}
where $\alpha$ is the fine structure constant. 
At the leading order, the hadronic tensor receives the contribution from Fig.\ref{SIDISFig1} where the gray box represents the density matrix in Eq.(\ref{DENM}). From Fig.\ref{SIDISFig1} the tensor is given by:  
\begin{eqnarray} 
  W^{\mu\nu} (P,q, k) = \frac{e_q^2}{(2\pi)^4}  \int d (x P^+)  (2\pi) \delta ((x P +q)^2)  \biggr ( \gamma^\mu \gamma\cdot ( x P+ q)  \gamma^\nu \biggr )_{ji} 
{ \mathcal M}_{F ij}(x, \xi, k_\perp) + \cdots, 
\end{eqnarray}  
where $\cdots$ are power-suppressed terms. With the fracture functions in Eq.(\ref{DEFQ}) we can derive the hadronic tensor at the order.  With the leptonic tensor at the leading order of QED, we 
have the differential cross-section in the target fragmentation region expressed with fracture functions as\cite{ABAK}: 
\begin{eqnarray} 
&&\frac{ d\sigma}{d x_B d y d \xi  d\psi d^2  k_{\perp} }      
\nonumber \\ 
&&= \frac{\alpha^2 }{ 2 (2\pi)^3 \xi y Q^2}  
\biggr [ (2-2y+ y^2) \biggr ( F_q (x_B,\xi, k_{\perp} ) + \frac{ \vert k_{\perp} \vert \vert  s_{\perp } \vert}   {m_A}
   \sin (\phi_s -\phi_h) 
      F_{qT} (x_B,\xi, k_{\perp})  \biggr )  
\nonumber\\
   &&\ \ \  + \lambda_e y (2-y)  \biggr ( 
    s_L \Delta  F_{q}  (x_B,\xi, k_{\perp} )  - \frac{ \vert k_{\perp} \vert \vert  s_{\perp } \vert}   {m_A}
    \cos (\phi_s- \phi_h) \Delta F_{q T} (x_B,\xi, k_{\perp} )  \biggr ) \biggr ], 
 \label{DIFF}    
\end{eqnarray} 
with 
\begin{equation} 
    s_\perp^\mu = \vert s_\perp \vert ( \cos \phi_s, \sin \phi_s), \quad k_\perp^\mu =\vert k_\perp \vert (
               \cos \phi_h, \sin \phi_h ).             
\end{equation} 
 We notice that the factorization in Eq.(\ref{DIFF}) holds in fact for the case 
with $k_\perp \ll Q$. In the case of $Q\gg k_\perp \gg \Lambda_{QCD}$, collinear factorization also holds. 
Therefore, fracture functions in this case can be matched to collinear parton distributions. The function 
$F_q$ and $\Delta F_{q}$ are matched to twist-2 parton distributions and fragmentation functions. 
The function 
$F_{qT}$ and $\Delta F_{qT}$ are matched to twist-3 parton distributions and twist-2 parton fragmentation functions.  It is interesting to note that there is no contribution involving twist-3 parton fragmentation functions 
because of helicity conservation. This will be explained later in detail.

\par\vskip10pt
 
\begin{figure}[hbt]
	\begin{center}
		\includegraphics[width=12cm]{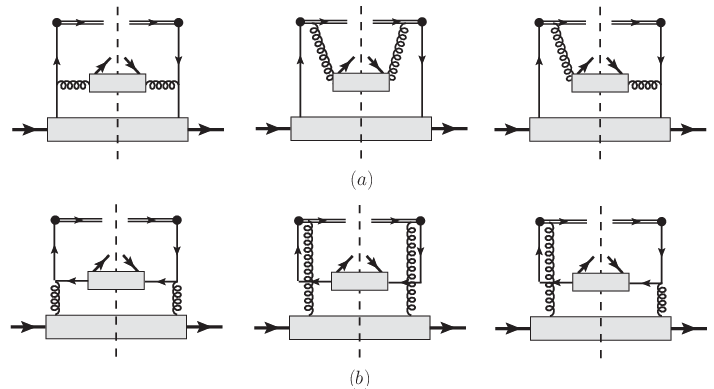}
	\end{center}
	\caption{Diagrams for the matching of quark fracture functions to parton distribution functions. (a): Contributions from quark parton distributions. (b): Contributions from gluon distributions. }
	\label{TreeFq}
\end{figure}

\par\vskip10pt
\noindent
{\bf 2.2.  Matching of $F_{q}$ and $\Delta F_{q}$}  
\par 

The fracture function $F_{q}$ and $\Delta F_{q}$ can be matched to twist-2 parton distributions and twist-2 parton fragmentation functions, in the case of $k_\perp \gg\Lambda_{QCD}$. 
These twist-2  quark- and gluon distributions can be defined as:
\begin{eqnarray} 
 && \int \frac{ dy}{2\pi}e^{-i y x P^+ } \langle h_A\vert \left (  \bar \psi (yn) 
  {\mathcal L}^\dagger _n (yn) \right )_\beta \left ( {\mathcal L} _n (0)  \psi (0) \right )_\alpha  \vert h_A\rangle
\nonumber\\  
   &&=   \frac{1}{2 N_c } \biggr [ q (x) \gamma^- + s_L \Delta q (x)\gamma_5 \gamma^- + h_1 (x) \gamma_5 \gamma\cdot s_\perp \gamma^-  \biggr ]_{\alpha\beta}  + \cdots, 
 \nonumber\\
 && \frac{1}{x P^+} \int \frac{ d\lambda}{2\pi}
e^{ - i x \lambda  P^+ }
    \langle h_A   \vert \left ( G^{+\mu } (\lambda n  ) {\mathcal L}_n^\dagger  (\lambda n) \right )^a  \left ( {\mathcal L}_n (0) G^{+\nu }(0) \right )^a \vert h_A  \rangle 
\nonumber\\    
     &&  = -\frac{1}{2} g_\perp^{\mu\nu}  g(x) -\frac{i}{2} \epsilon_\perp^{\mu\nu} s_L \Delta g(x) + \cdots. 
\label{TW2PDF} 
\end{eqnarray}
In the above,  $q(x)$ is the spin-averaged quark distribution function, 
$\Delta q $ is the longitudinally polarized quark distribution, $h_1$ is the quark transversity distribution. 
$\cdots$ stand for twist-3 or higher twists.  At twist-2 there are two gluon distributions.  $g(x)$ is  the spin-averaged gluon distribution function, 
$\Delta g $ is the longitudinally polarized gluon distribution. 
At twist-2 the relevant parton fragmentation functions of an unpolarized hadron are defined as:
\begin{eqnarray} 
   d_{\bar q} (z) &=& \frac{ z}{2 N_c} \int\frac{ d\lambda}{4\pi} e^{- i \lambda k^+/z} \langle 0\vert 
     \bar \psi (0) {\mathcal L}^\dagger _n (0) \gamma^+ \vert h, X\rangle \langle X, h\vert {\mathcal L}_n (\lambda n) \psi(\lambda n) \vert 0 \rangle,   
\nonumber\\
   d_g (z)&=& \frac{ z} { (N_c^2-1) k^+ } \int\frac{ d\lambda}{4\pi} e^{- i \lambda k^+/z} \langle 0\vert 
     G^{a,+\mu}(0)  {\mathcal L}^\dagger _n (0)  \vert h, X\rangle \langle X, h\vert {\mathcal L}_n (\lambda n) G^{a,+}_{\quad\  \mu} (\lambda n)   \vert 0 \rangle,   
\end{eqnarray}
where $d_{\bar q}$ is the anti-quark fragmentation function, $d_g$ is the gluon fragmentation function. 
The observed hadron in the final state is with the momentum $k^{\mu} = (k^+,0,0,0)$, where its mass is neglected.

With the twist-2 parton distributions and fragmentation functions, one can easily calculate the fracture functions $F_q$ and $\Delta F_q$ at tree-level, where nonperturbative effects are represented by these distributions and 
functions.  The contributions to $F_q$ and $\Delta F_q$ are given by Fig.2.  Calculations of these diagrams are rather standard. Therefore, we give our twist-2 results directly. More technical details will be explained in subsequent sections. From Fig.2 we obtain the fracture functions $F_q$ and  $\Delta F_q$: 
\begin{eqnarray} 
  F_q (x,\xi,k_\perp) &=&  g_s^2  \frac{1}{k_\perp^2} \int \frac{ dz}{z^2} \biggr [ 2 C_F d_g (z)
     q(y) \frac{z^2}{y^2} (x^2 + y^2) + d_{\bar q} (z) g (y)  \frac{  \xi }{z y^3} ( z^2 x^2 + \xi^2 ) \biggr ], 
\nonumber\\
  \Delta F_q (x,\xi,k_\perp) &=&  g_s^2  \frac{1}{k_\perp^2} \int \frac{  dz}{z^2} \biggr [ 2 C_F  d_g (z)
     \Delta q(y) \frac{z^2}{y^2} (x^2 + y^2) +d_{\bar q}  (z)
     \Delta g(y) \frac{\xi }{ y^2} (xz- \xi) \biggr ] ,
\label{FQDFQ}        
\end{eqnarray}     
with 
\begin{equation} 
    y = x+ \frac{\xi}{z}. 
\end{equation}     
We notice that for $k_\perp\gg \Lambda_{QCD}$, the twist-2 fracture functions behave like $1/k_\perp^2$.

\par 
The other two fracture functions $F_{qT}$ and $\Delta F_{qT}$ are for the initial hadron which is transversely polarized. At first look, the twist-2 transversity distribution $h_1$, which is chirality-odd,  can give contributions to these functions from Fig.2, where twist-3 parton fragmentation functions are involved. 
At the order we consider, helicity or chirality is conserved in the amplitude of the upper part in Fig.1.  
Therefore, contributions involve chirality-odd parton distributions 
are absent. 
Hence, the two fracture functions only receive contributions where twist-3 parton distributions and twist-2 parton fragmentation functions are involved. Before we turn to the matching of these two functions, we give definitions of twist-3 parton distributions in the next subsection. 

\par\vskip10pt
\noindent
{\bf 2.3.   Definitions of Twist-3 Parton Distributions}  
\par  
\par 
From the quark density matrix  one can define a set of  three twist-3 parton distributions with one transverse 
derivative\cite{CMZ}:
\begin{eqnarray} 
  q_T (x) s_\perp^\mu  &=&  P^+ \int \frac{d\lambda}{ 4\pi } e^{- i x\lambda  P^+} \langle h_A  \vert \bar \psi(\lambda n) {\mathcal L}^\dagger_n(\lambda n)  \gamma_\perp^\mu \gamma_5  
  {\mathcal L}_n(0) \psi(0) \vert h_A  \rangle, 
\nonumber\\  
  -i q_\partial (x) s_\perp^\mu &=&    \int \frac{d\lambda}{ 4\pi } e^{- i x\lambda  P^+} \langle h_A  \vert \bar \psi(\lambda n) {\mathcal L}^\dagger_n(\lambda n)  \gamma^+ \gamma_5 \partial_\perp^{\mu}   
  \left ( {\mathcal L}_n  \psi \right ) (0) \vert h_A \rangle,
\nonumber\\  
  -i q_\partial' (x) \tilde s_\perp^\mu &=&    \int \frac{d\lambda}{ 4\pi } e^{- i x\lambda  P^+} \langle h_A  \vert \bar \psi(\lambda n) {\mathcal L}^\dagger_n(\lambda n)  \gamma^+ \partial_\perp^{\mu}   
  \left ( {\mathcal L}_n  \psi \right ) (0) \vert h_A \rangle,    
\label{tw31}  
\end{eqnarray}
with $\tilde s^\mu =\epsilon_\perp^{\mu\nu} s_{\perp\nu}$. 
 The three distributions are real.
\par 
Another set of  twist-3 distributions can be defined with a pair of quark fields with one gluon field strength operator. They are given by the matrix:
\begin{eqnarray} 
{\mathcal M}^\mu_{F\alpha\beta} (x_1,x_2)  =  g_s \int\frac{d y_1 dy_2}{2\pi} e^{-iy_1x_1 P^+ -i y_2 (x_2-x_1) P^+} 
   \langle h_A \vert \bar \psi_\beta (y_1 n) G^{+\mu}(y_2 n)  \psi_\alpha (0) \vert h_A \rangle,
\end{eqnarray}
where we have suppressed the gauge links for short notations. 
The matrix can be decomposed into
\begin{eqnarray} 
  {\mathcal M}^\mu_{F} (x_1,x_2) &=& \frac{1}{2} \biggr [ T_F (x_1,x_2) \tilde s^\mu_\perp \gamma^- 
     + T_{\Delta} (x_1,x_2) is^\mu_\perp \gamma_5\gamma^- \biggr ] 
\nonumber\\
    &&  + \frac{1}{4} \biggr [ s_L \tilde T_{\Delta ,F} (x_1,x_2) i \gamma_5 \gamma^\mu_\perp \gamma^- 
           + \tilde T_F  (x_1,x_2) i\gamma^\mu_\perp \gamma^ -\biggr ]. 
\label{tw3}             
\end{eqnarray} 
Properties of these twist-3 parton distributions have been studied in \cite{EFTE,QiuSt,JKT,ZYL,BD}. 
The first two have been introduced in the study of single transverse-spin asymmetries. The last two are chirality-odd. They will not give contributions in the matching because of helicity conservation. 
The chirality-even parton distributions satisfy: 
\begin{eqnarray} 
     T_F (x_1,x_2) =  T_F(x_2,x_1), \quad \quad  T_{\Delta} (x_1,x_2) = - T_{\Delta} (x_2,x_1).             
\label{TW3P}
\end{eqnarray}
Corresponding to the two distributions in Eq.(\ref{tw3})  
one can define additionally two twist-3 distributions by replacing the field strength tensor $g_s G^{+\mu}(x)$ 
with $P^+ D^\mu_\perp(x)$, where $D^\mu(x)$ is given by $D^\mu (x)=\partial_\mu + ig_s G^\mu (x)$. These two functions 
will not appear in our calculation. In fact they can be expressed with the distributions given 
in Eqs.(\ref{tw31},\ref{tw3}) as shown in \cite{EKT}.

The three twist-3 distributions  in Eq.(\ref{tw31}) and the two twist-3 chirality-even distributions in Eq.(\ref{tw3}) are not independent. One can show:   
\begin{equation} 
\frac{1}{2\pi}\int d x_1  P\frac{ 1 }{x_1-x_2}  \biggr [ T_F (x_1,x_2)  - T_\Delta (x_1,x_2) \biggr ] = -x_2 q_T (x_2) + q_\partial (x_2), \quad
   T_F (x,x) = - 2 q'_\partial (x),   
\label{RL1} 
\end{equation}   
where $P$ stands for the principle-value prescription. The first relation has been derived in \cite{JiG2}.  The second relation  
is derived in \cite{CMZ}.  It should be emphasized that 
the second relation is for SIDIS, for $q_{\partial}'$ defined with future pointing gauge links.  The distribution $q'_\partial (x)$ in Drell-Yan processes is defined with gauge links pointing to the past. 
With the symmetries of time-reversal and parity one can show that there is a sign-difference between the two distributions. 

There are twist-3 gluon distributions defined only with gluon fields. One can first define three density matrices:
\begin{eqnarray}
M_\partial^{\mu\nu\rho} (x) &=&  \frac{1}{x P^+} \int \frac{ d\lambda}{2\pi}
e^{ - i x \lambda  P^+ }
    \langle h_A   \vert \left ( G^{+\mu } (\lambda n  ) {\mathcal L}_n^\dagger  (\lambda n) \right )^a 
    \partial_\perp^\rho  \left ( {\mathcal L}_n (0) G^{+\nu }(0) \right )^a \vert h_A   \rangle
\nonumber\\
M_D^{\mu\nu\rho} (x_1,x_2) &=&    \int \frac{d\lambda_1}{2\pi} \frac{ d\lambda_2}{2\pi} e^{-i\lambda_1 x_1 P^+ -  i\lambda_2 (x_2-x_1)P^+}
 \nonumber\\   
   &&    
   \langle h_A \vert  \biggr ( G^{+\mu }(\lambda_1n) {\mathcal L}(\lambda_1n )  \biggr )_a 
       \biggr (  {\mathcal L} (\lambda_2 n)  D_\perp^\rho (\lambda_2 n ) {\mathcal L}^\dagger  (\lambda_2 n) \biggr )_{ab}   \biggr ( {\mathcal L} (0 )   G^{+\nu} (0  )  \biggr )_b  \vert h_A  \rangle,
\nonumber\\
M_F^{\mu\nu\rho } (x_1,x_2 ) &=&  -\frac{g_s}  {P^+} \int \frac{d\lambda_1}{2\pi} \frac{ d\lambda_2}{2\pi} e^{-i\lambda_1 x_1 P^+ - i\lambda_2 (x_2-x_1)P^+}
\nonumber\\   
   &&    \langle h_A \vert  \biggr ( G^{+\mu}(\lambda_1n) {\mathcal L}(\lambda_1n )  \biggr )_a 
       \biggr (  {\mathcal L} (\lambda_2 n)  G^{+\rho}  (\lambda_2 n ) {\mathcal L}^\dagger  (\lambda_2 n) \biggr )_{ab}   \biggr ( {\mathcal L} (0 )   G^{+\nu} (0  )  \biggr )_b  \vert h_A  \rangle,
\label{TW3GM}        
\end{eqnarray}
where the field strength tensor $G^{+\rho}  (\lambda_2 n )$ in the last line and  the covariant derivative $D^\rho (\lambda_2 n)$ are in the adjoint representation.  The three matrices are not independent. They are  related through the relation:
\begin{equation} 
M_D^{\mu\nu\rho} (x_1,x_2) = x_1 \delta (x_1-x_2) M^{\mu\nu\rho}_\partial (x_1) - \frac{1}{x_2 -x_1 + i\varepsilon} M_F^{\mu\nu\rho} (x_1,x_2). 
\label{RL3G} 
\end{equation} 
Besides the three matrices at twist-3 one can also find an additional twist-3 matrices conveniently in 
the light-cone gauge $G^+=0$. In this gauge, one can introduce 
\begin{eqnarray} 
&& \frac{i^3 g_s}{P^+}  \int \frac{d\lambda_1}{2\pi }\frac{d\lambda_2}{2\pi} 
   e^{i\lambda_1 x_1 P^+ + i\lambda_2 (x_2-x_1)P^+} \langle h_A  \vert G^{a,+\alpha} 
   (\lambda_1 n)  G^{c,+\gamma}(\lambda_2 n) G^{b,+\beta} (0) \vert h_A  \rangle 
\nonumber\\
   && = \frac{N_c}{(N_c^2-1)(N_c^2-4)} d^{abc} O^{\alpha\beta\gamma}(x_1,x_2) - \frac{i }{N_c(N_c^2-1) } f^{abc} N^{\alpha\beta\gamma}(x_1,x_2), 
\label{NO3G}    
\end{eqnarray} 
where all indices $\alpha,\beta$ and $\gamma$ are transverse. One can identify that:
\begin{equation} 
M_F^{\mu\nu\rho } (x_1,x_2 )  =  i N^{\mu\nu\rho} (-x_1, - x_2). 
\end{equation} 
Therefore, there are two twist-3 matrices built with three gluon field strength tensor operators. From Bose-symmetry and covariance                
the two tensors take the form \cite{Ji3G,KTY,BKTY}: 
\begin{eqnarray}
O^{\alpha\beta\gamma}(x_1,x_2) &=& -2 i \biggr [ O(x_1,x_2) g^{\alpha\beta}_\perp \tilde s_\perp^\gamma + 
  O(x_2,x_2-x_1) g^{\beta\gamma}_\perp \tilde s_\perp^\alpha + O(x_1,x_1-x_2) g^{\gamma\alpha}_\perp \tilde s_\perp^\beta \biggr ], 
\nonumber\\
N^{\alpha\beta\gamma}(x_1,x_2) &=& -2 i \biggr [ N(x_1,x_2) g^{\alpha\beta}_\perp \tilde s_\perp^\gamma - 
  N(x_2,x_2-x_1) g^{\beta\gamma}_\perp \tilde s_\perp^\alpha - N(x_1,x_1-x_2) g^{\gamma\alpha}_\perp \tilde s_\perp^\beta \biggr ],
\label{NOG3}           
\end{eqnarray}
with the properties of the functions $O$ and $N$ 
\begin{eqnarray} 
    && O(x_1,x_2) =O(x_2,x_1),\quad O(x_1,x_2) = O(-x_1,-x_2), 
\nonumber\\ 
   && N(x_1,x_2) = N(x_2,x_1), \quad
N(x_1,x_2) = -N(-x_1,-x_2). 
\end{eqnarray}
The matrix $M_\partial ^{\mu\nu\rho} (x)$ can be parametrized with constraints of symmetries as: 
\begin{eqnarray} 
M_\partial^{\mu\nu\rho} (x) = i g_{\perp}^{\mu\nu} \tilde s_\perp^\rho g_{\partial}(x) 
 + i  \biggr ( g_{\perp}^{\mu\rho} 
\tilde s_\perp^\nu + g_\perp^{\nu\rho} \tilde s_\perp^{\mu} \biggr ) g_{\partial}' (x)
\label{MPGG} 
\end{eqnarray} 
with 
\begin{eqnarray} 
\quad g_\partial(x) = g_\partial (-x), \quad \quad g_\partial '(x) = g_\partial '(-x).   
\end{eqnarray} 
The two functions $g_\partial$ and $g_{\partial}'$ are real.              
From the relation in Eq.(\ref{RL3G}) one has:
\begin{equation} 
   x g_\partial (x) = 2\pi N(x,x), \quad xg_\partial' (x) = - 2\pi N(x,0). 
\end{equation}  
Therefore, all twist-3 gluon distributions are determined by $N(x_1,x_2)$ and $O(x_1,x_2)$. 

In our notations,  
all twist-3 parton distributions have the dimension $1$ in mass and are proportional 
to $\Lambda_{QCD}$.  

 \par\vskip20pt
\noindent 
{\bf 3. Matching of $F_{qT}$} 

$F_{qT}$ describes the single transverse-spin asymmetry of SIDIS in target fragmentation region.  It is well-known that such an asymmetry is a T-odd effect which requires nonzero absorptive parts in the scattering amplitude. In the case of $k_\perp \gg \Lambda_{QCD}$ the asymmetry is a twist-3 effect in the collinear factorization\cite{ EFTE,QiuSt}. In this case, $F_{qT}$ can be matched with twist-3 parton distributions and twist-2 parton fragmentation functions.  Because of helicity conservation, the contribution involving the twist-2 quark distribution $h_1$ with twist-3 parton fragmentation functions is absent as discussed before.

\begin{figure}[hbt]
	\begin{center}
		\includegraphics[width=8cm]{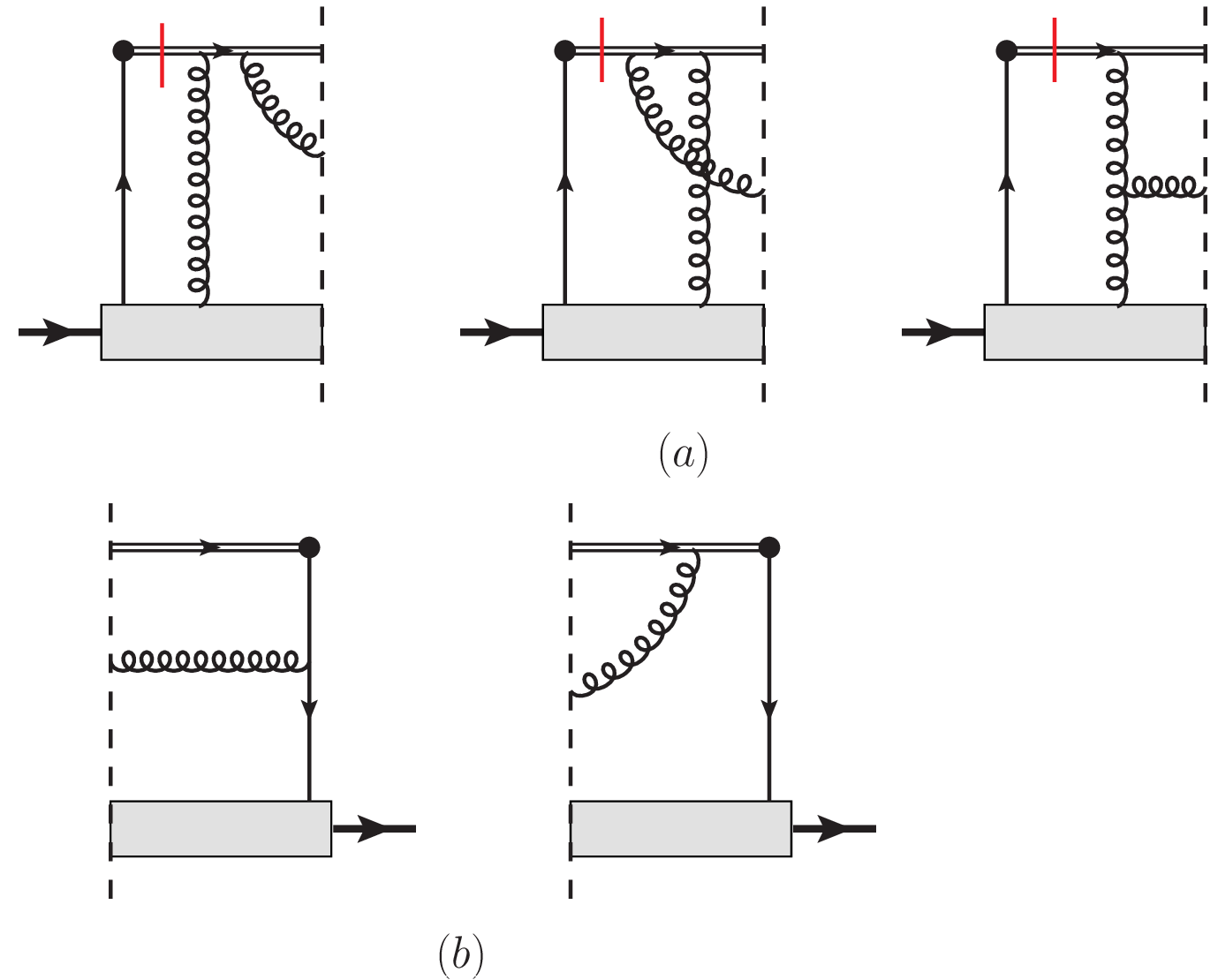}
	\end{center}
	\caption{Diagrams for hard-pole contribution.  (a) The left part of diagrams, where there is one gluon attached to the lower bubble.   (b) The right-part of diagrams. The contribution is given by the interference between the left- and right diagrams.   }
	\label{HDP}
\end{figure} 
 \par
\begin{figure}[hbt]
	\begin{center}
		\includegraphics[width=11cm]{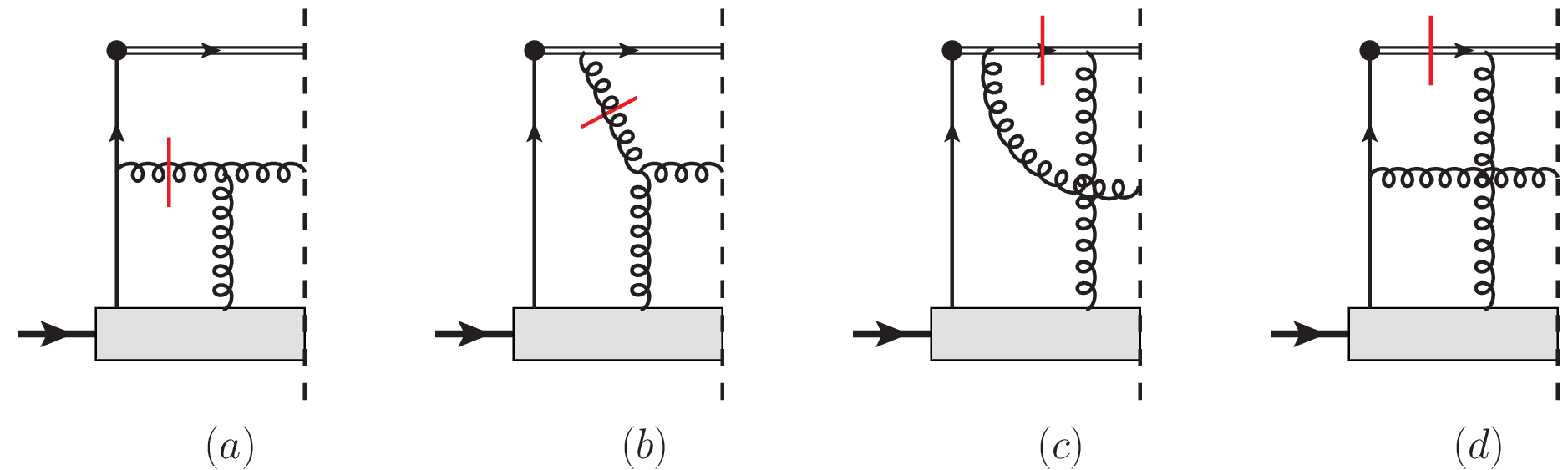}
	\end{center}
	\caption{The left parts of diagrams for soft-gluon-pole contribution.   }
	\label{SGP}
\end{figure}

\begin{figure}[hbt]
	\begin{center}
		\includegraphics[width=7cm]{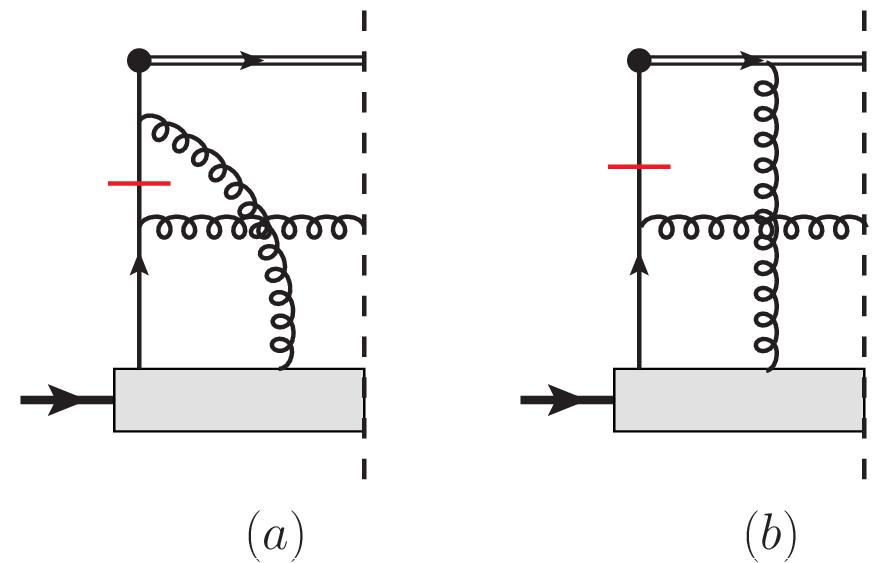}
	\end{center}
	\caption{The left parts of diagrams  for soft-fermion-pole  contribution.   }
	\label{SFP}
\end{figure}

 
At the order considered, there is no contribution from Fig.\ref{TreeFq} because of the absence of the required absorptive part.  The contributions are given by Fig.\ref{HDP}, Fig.\ref{SGP} and Fig.\ref{SFP}, 
where the diagrams are given in terms of amplitude. The contributions are obtained from the interference of  the amplitudes of Fig.\ref{HDP}a, Fig.\ref{SGP} and Fig.\ref{SFP} with those of Fig.\ref{HDP}b.  
There is one propagator in Fig.\ref{HDP}a, Fig.\ref{SGP} and Fig.\ref{SFP} with a short bar. This implies that only the absorptive part of the propagator is taken into account. This gives the required absorptive parts in the amplitudes. The contributions are classified according to the parton momenta. The hard-pole contribution is from Fig.\ref{HDP} where the gluon parton carries nonzero momentum. The soft-gluon contribution is from Fig.\ref{SGP} where the gluon parton carries zero momentum. The soft-quark contribution is from Fig.\ref{SFP} where the quark parton is with zero momentum. 

All contributions from these diagrams involve only the twist-2 gluon fragmentation function. 
The general structure of each type of the contributions from these diagrams can be 
written in the form
\begin{eqnarray} 
  \frac{1}{m_A}F_{qT} \biggr\vert _L = 
\int \frac{ d z }{z^2} d_g(z)  d^4 k_A d^4 k_1    
   {\mathcal H}_{L, kl}^{a,\rho}(k_A, k_1) {\mathcal M}^{a}_{\rho,lk}(k_A, k_1)   , 
\end{eqnarray}
with the quark-gluon correlator:     
\begin{eqnarray}             
 {\mathcal M}^{a,\rho} ( k_A, k_1) = g_s\int \frac{d^4 \xi_1   d^4\xi_2 }{(2\pi)^8 } 
     e^{  -i  \xi_1\cdot  k_A +i \xi_2 \cdot  k_1 } 
       \langle h_A \vert \bar q ( 0  ) G^{a,\rho} (\xi_2 ) q ( \xi_1 ) \vert h_A \rangle .  
\label{QGC}                 
\end{eqnarray}
$d_g$ is the gluon fragmentation function.  ${\mathcal H}_L^{a,\rho}$ is the sum of upper parts of all diagrams in which the gluon line from the correlator in the left 
side of the cut, or in the left part of diagrams. The quark in the right part of the diagrams is with the momentum $k_A$, the gluon is with $k_1$ and the quark in the left part 
is with the momentum $k_A -  k_1$.   

The quark-gluon density matrix can be written in the form: 
\begin{eqnarray}
{\mathcal M}^{a,\rho} ( k_A, k_1) &=& \frac{1}{N_c^2-1} T^a \biggr [ \biggr ( M^\rho (k_A, k_1)  \gamma^- 
  + M_A^\rho( k_A,  k_1) i \gamma_5 \gamma^- \biggr )
 \nonumber\\ 
     &&    + \biggr ( M^{\rho \mu}_\perp  
     (k_A, k_1)  \gamma_{\perp \mu} 
  + M_{A\perp} ^{\rho\mu} ( k_A,  k_1) i \gamma_5 \gamma_{\perp\mu} \biggr ) \biggr ] 
  +\cdots,
\end{eqnarray} 
where $\cdots$ denote the terms beyond twist-3. The quark-gluon matrix elements are 
\begin{eqnarray} 
 M^\rho (k_A, k_1)  &=& g_s\int \frac{d^4 \xi_1   d^4\xi_2 }{2 (2\pi)^8} 
     e^{ - i  \xi_1\cdot  k_A +i \xi_2 \cdot  k_1 } 
       \langle h_A \vert \bar q (\xi_1 )  \gamma^+ G^{\rho} (\xi_2 ) q ( 0 ) \vert h_A \rangle,    
\nonumber\\
M_A^\rho( k_A,  k_1) &=&   -i g_s\int \frac{d^4 \xi_1   d^4\xi_2 }{2 (2\pi)^8}  
     e^{ - i  \xi_1\cdot  k_A +i \xi_2 \cdot  k_1 } 
       \langle h_A \vert \bar q (\xi_1 )  \gamma^+ \gamma_5 G^{\rho} (\xi_2 ) q ( 0 ) \vert h_A \rangle,   
\nonumber\\
 M^{\rho\mu}_\perp (k_A, k_1)  &=& g_s\int \frac{d^4 \xi_1   d^4\xi_2 }{2 (2\pi)^8}  
     e^{ - i  \xi_1\cdot  k_A +i \xi_2 \cdot  k_1 } 
       \langle h_A \vert \bar q (\xi_1 )  \gamma_\perp^\mu G^{\rho} (\xi_2 ) q ( 0 ) \vert h_A \rangle ,    
\nonumber\\
M_{A\perp} ^{\rho\mu} ( k_A,  k_1) &=&   -i g_s\int \frac{d^4 \xi_1   d^4\xi_2 }{2 (2\pi)^8}  
     e^{ - i  \xi_1\cdot  k_A +i \xi_2 \cdot  k_1 } 
       \langle h_A \vert \bar q (\xi_1 )  \gamma_\perp^\mu  \gamma_5 G^{\rho} (\xi_2 ) q ( 0 ) \vert h_A \rangle .  
 \label{A2}   
\end{eqnarray}
The power counting for the parton momenta and the quark-gluon matrix elements are:
\begin{eqnarray} 
 &&  k_{A}^\mu \sim (1,\lambda^2,\lambda,\lambda),\quad k_1^\mu \sim (1,\lambda^2,\lambda,\lambda),
\nonumber\\
   &&  M^{\rho\mu}_\perp \sim M_{A\perp} ^{\rho\mu}  \sim  \lambda (1,\lambda^2, \lambda,\lambda), \quad M^\rho \sim M_A^\rho \sim (1,\lambda^2, \lambda,\lambda). 
\label{PWC} 
 \end{eqnarray} 
In the case of $F_{qT}$,  the matrix elements $M^{\rho\mu}_\perp$ and  $M_{A\perp} ^{\rho\mu}$  
contribute at twist-3 only with $\rho =+$, where one can neglect all transverse- and $-$-components of parton momenta. 
 For $\rho\neq +$, the contributions from these two matrix elements are beyond twist-3.

We take the contributions from $M^\rho(k_A,k_1)$ in Eq.(\ref{A2}) to discuss  the collinear expansion and the gauge invariance of the results.  The discussion in the case of $M_A^\rho(k_A,k_1)$  is similar.  
To obtain twist-3 contributions we need to expand the upper parts around the parton momenta:
 \begin{equation}
 \hat k_A^\mu =(k_A^+,0,0,0), \quad \hat k_1^\mu= (k_1^+,0,0,0).
 \label{PC}  
\end{equation} 
The upper part of the contributions involving $M^\rho(k_A,k)$ is given by:
\begin{equation} 
   H_L^\mu (k_A, k_1) = \frac{1}{ (N_c^2-1) } {\rm Tr }  \left [ T^a \gamma^- {\mathcal H}_L^{a,\mu}(k_A,k_1) 
   \right ]. 
\end{equation}    
The collinear expansion is:
\begin{equation} 
   H_L^ \mu   (k_A,k_1)  = H_L^\mu  (\hat k_A, \hat k_1) + k_{1\perp} ^\alpha \frac{\partial H_L^\mu }{\partial k_{1\perp} ^\alpha }(\hat k_A,\hat k_1) + k_{A\perp}^\alpha \frac{\partial H_L^\mu }{\partial k_{A\perp}^\alpha }(\hat k_A,\hat k_1) + \cdots , 
\end{equation} 
where terms represented by $\cdots$ give the contributions beyond twist-3.  With the expansion, the contribution at the leading order of $\lambda$ can be written 
in the form: 
 \begin{eqnarray} 
 \frac{1}{m_A}F_{qT}  \biggr\vert_L &=&   g_s  \int  d k_A^+  d k_1^+ \frac{ d z}{z^2} d_g (z)  \int \frac{d\lambda_1 d\lambda_2}{2 (2\pi)^2} e^{-i\lambda_1 k_A^+ +i \lambda_2 k^+_1}  \biggr \{  
\nonumber\\ 
   &&  i   \frac{\partial H_L^-}{\partial k_{1\perp} ^\alpha }(\hat k_A,\hat k_1)
    \langle h_A  \vert \bar q (\lambda_1  ) \gamma^+  \biggr [ \partial_\perp^\alpha G^+ 
    - \partial^+ G^{\alpha}_\perp \biggr ](\lambda_2 n ) q(0)  \vert h_A  \rangle 
\nonumber\\
  &&    -i  \frac{\partial H_L^-}{\partial k_{A\perp}^\alpha }(\hat k_A,\hat k_1)  
    \langle h_A \vert  \left ( \partial_{\perp}^\alpha \bar q  \right ) (\lambda_1  ) \gamma^+   G^ +(\lambda_2 n)  q(0)  \vert h_A  \rangle  
 \nonumber\\
 && + \biggr [  k_1^+ \frac{\partial H_L^-}{\partial k_{1\perp}^\alpha }(\hat k_A,\hat k_1)
    + H_{L\perp\alpha} (\hat k_A, \hat k_1) \biggr ] 
    \langle h_A \vert \bar q (\lambda_1  ) \gamma^+   G_{\perp}^\alpha (\lambda_2 n)  q(0)  \vert h_A  \rangle    
    \biggr \} , 
\label{GI1} 
  \end{eqnarray}  
where  contributions beyond the leading power of $\lambda$ are neglected. 

\par 
 It is clear that the first term  in Eq.(\ref{GI1}) is gauge-invariant at the considered order of $g_s$. The second- and third terms in Eq.(\ref{GI1}) are not gauge-invariant.  However,  because of the additional cut, i.e., 
that  the propagator with a short bar represents an on-shell particle,  one can find the following Ward identity
and the identity with the momenta $\hat k_{1,A}$:
\begin{equation} 
     k_{1\rho} H_L^\rho (k_A,k_1) =0, \quad k_1^+ H_L^- (\hat k_A, \hat k_1) =0 . 
\label{WID}         
\end{equation} 
With the first identity, one can show that in the hard-pole- and soft-quark contributions the second- and third
gauge-variant terms are zero.  In the case of the soft-gluon pole contribution, because of that $k_1^+=0$ one can not use the argument of the identity. But, we find by adding the contribution from complex conjugated diagrams that the final result is gauge invariant. The situation here is similar to the analysis of twist-3 contribution of SIDIS in \cite{EKT2}.  For the contributions from $M_A^\rho(k_A,k_1)$ 
the results are similar except the soft-gluon-pole contribution which is nonzero and gauge-variant.   

For the contributions from the matrix elements $M^{\rho\mu}_\perp$ and  $M_{A\perp} ^{\rho\mu}$  with $\rho=+$ there are similar identities to those in Eq.(\ref{WID}). In the hard-pole- and soft-quark contributions only the momentum component $k_1^+$ and $k_A^+$  are not zero. Therefore, there are no hard-pole- and soft-quark contributions from these two matrix elements because of an identity similar to the second one in Eq.(\ref{WID}). 
For soft-gluon-contributions, the argument can not be used because of that $k_1^+=0$. 
With explicit calculations we find that the soft-gluon-pole contribution from $M^{+\mu}_\perp$ 
is zero if we add  the contribution from complex conjugated diagrams. The soft-gluon-pole contribution from 
 $M_{A\perp} ^{+\mu}$ and that from $M_A^\rho$  are not zero. Both contributions can not be written in a gauge-invariant form.  We notice that in the matrix element $M_{A\perp} ^{+\mu}$ one of the quark fields must be the $-$-component with the decomposition discussed 
 in the next section. With equation of motion one can show that the sum of soft-gluon-pole contributions 
 from $M_{A\perp} ^{+\mu}$ and $M_A^\rho$ can be written in a gauge invariant form, i.e., in terms of $T_F$. Therefore,  the soft-gluon-pole contributions come from all matrix elements in Eq.(\ref{A2}) except 
 $M^{\rho\mu}_\perp$ with $\rho=+$. The calculation is tedious but straightforward. Hence, we will give our results of this section directly without giving the details of our calculations.

The hard-pole contribution is from Fig.\ref{HDP}. The result is:
\begin{eqnarray} 
 \frac{1}{m_A} F_{qT} (x,\xi,k_\perp) \biggr\vert_{HP}  =g_s^2 \frac{N_c}{ (k_\perp^2)^2} \int  \frac{ dz }{z^2} d_g (z) \frac{z^2}{y} \biggr ( 
    \xi T_{\Delta }( y,x) - (\xi + 2 xz ) T_F (y,x) \biggr ). 
\label{FQTHDP}     
\end{eqnarray}  
The soft-gluon-pole contribution is from Fig.\ref{SGP}. The result is   
\begin{eqnarray} 
   \frac{1}{m_A} F_{qT} (x,\xi,k_\perp) \biggr\vert_{SGP}  &=&\frac{ g_s^2 N_c}{ (k_\perp^2)^2} \int \frac{ dz} {z^2} d_g (z) \frac{1}{y^3} \biggr [  z^3 (y^3 +3 x^2 y -2 x^3)  T_F (y,y)  
\nonumber\\
    && -  y \xi z^2(y^2 + x^2)  \frac{ \partial T_F (y,y) }{\partial  y}  \biggr ]    . 
\label{FQTSGP}      
\end{eqnarray}
There is no soft-gluon contribution from $ T_{\Delta }$ because of that $ T_{\Delta }(y,y)=0$. The soft-quark contribution is from Fig.\ref{SFP}. The result is  
\begin{eqnarray}
   \frac{1}{m_A} F_{qT} (x,\xi,k_\perp) \biggr\vert_{SFP}= g_s^2 \frac{1}{N_c} \frac{1}{( k_\perp^2)^2} \int \frac{dz}{z^2} d_g (z) \frac{ x\xi z}{y^3} 
       \biggr ( (xz-\xi) T_F (y,0) - (\xi + xz ) T_\Delta (y,0) \biggr ). 
\label{FQTSFP}        
\end{eqnarray}     
In these results $y$ is given by $y= x+ \xi/z$.

\begin{figure}[hbt]
	\begin{center}
		\includegraphics[width=11cm]{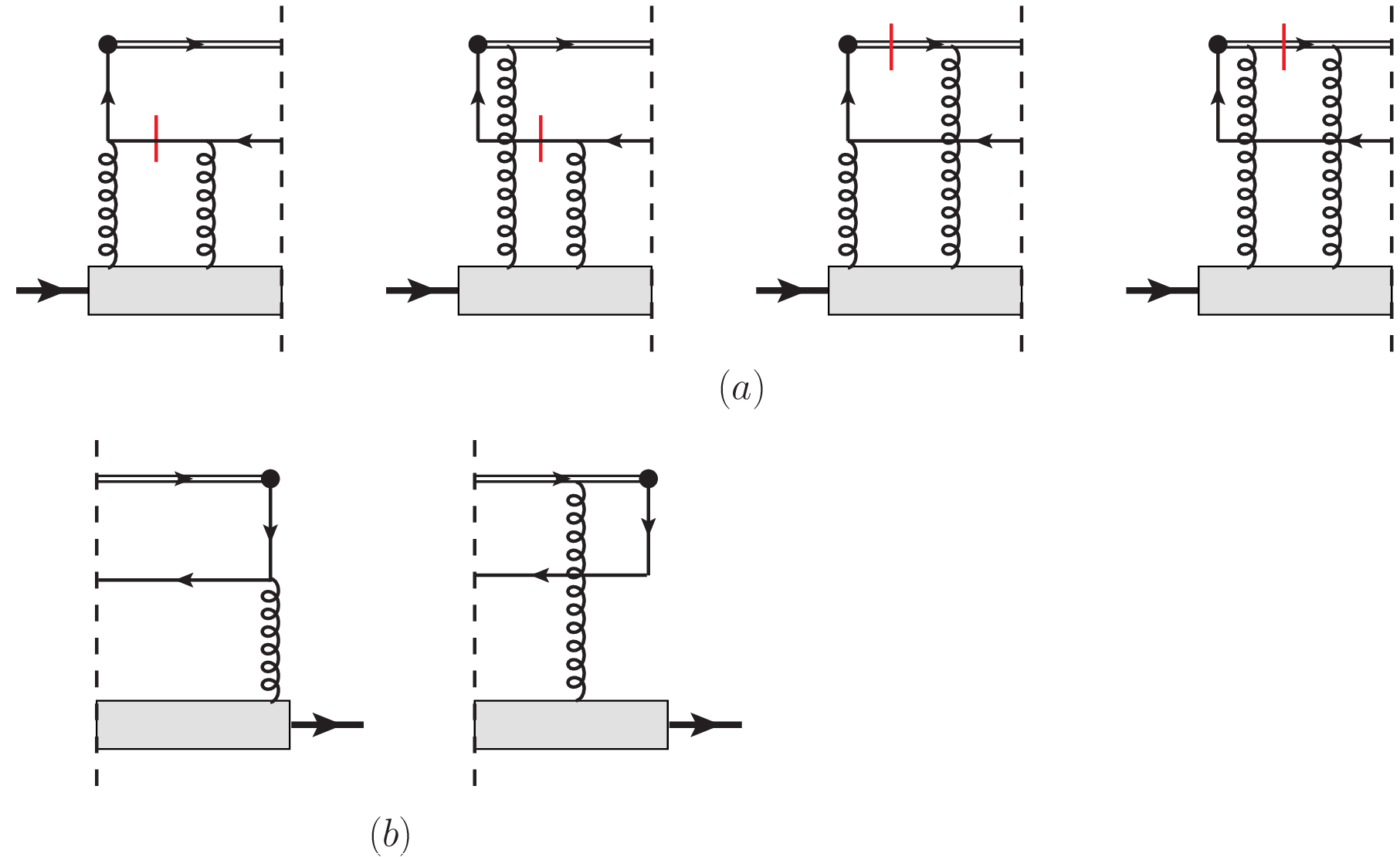}
	\end{center}
	\caption{Diagrams  for twist-3 gluonic contribution.   }
	\label{G3}
\end{figure}

Besides the twist-3 contributions from quark-gluon correlators, there are contributions from purely gluonic distributions at twist-3.  Since an additional cut or an absorptive part is required, there is no contribution from Fig.2b. The contributions are only from Fig.\ref{G3}, 
where one quark propagator or eikonal propagator is with a short bar. 

After completing the collinear expansion related to the produced 
hadron,  the contribution from Fig.\ref{G3} can be written in the form: 
\begin{eqnarray} 
\frac{1}{m_A} F_{qT} (x,\xi, k_\perp) \biggr\vert_L  &=& \int \frac{d z}{z^2} d_{\bar q} (z) 
 d^4 k_1 d^4 k_2  H^{abc}_{L \mu_1\mu_2 \mu_3} 
     (k_1,k_2) 
\nonumber\\
   && \int \frac{ d^4 x_1 d^4 x_2}{ (2\pi)^8} e^{ i x_1\cdot k_1 + ix_2\cdot k_2} 
   \langle h_A  \vert  G^{c,\mu_3} (0) G^{b,\mu_2} (x_2) G^{a,\mu_1}(x_1) \vert h_A \rangle. 
\end{eqnarray}           
This contribution is associated with the fragmentation function of an antiquark into the observed hadron.
$H^{abc}_{L \mu_1\mu_2 \mu_3} (k_1,k_2)$  is the perturbative part which is the sum of the upper parts of Fig.\ref{G3}.  In Fig.\ref{G3}a the left gluon line carries the momentum $k_1$ and the right one carries $k_2$.  In Fig.\ref{G3}b, the gluon line carries the momentum $k_3$.   Because of the cut and that the propagator with a short bar represents an on-shell particle, one can find the following identities: 
\begin{equation} 
  k_1^{\mu_1} H^{abc}_{L \mu_1\mu_2 \mu_3} (k_1,k_2) =0, \quad  k_3^{\mu_3} H^{abc}_{L \mu_1\mu_2 \mu_3} (k_1,k_2) =0, \quad k_3 = -(k_1 +k_2). 
\end{equation}
The contributions from the last two diagrams  in Fig.\ref{G3}a are proportional to $\delta(k_2^+)$ because of the cut on the eikonal propagator. With this fact we can find the identity:  
\begin{equation} 
  k_2^{\mu_2} H^{abc}_{L \mu_1\mu_2 \mu_3} (k_1,k_2) =0. 
  \end{equation}
 However, this identity is useless in the case of soft-gluon-pole contributions here. This is similar to the case discussed after Eq.(41). We find the leading contribution from Fig.\ref{G3} and their complex conjugated diagrams can be written in a gauge-invariant form. Other contributions are beyond twist-3. We have the result 
with the twist-3 gluon distributions: 
\begin{eqnarray}
  && \frac{1}{m_A}F_{qT} (x,\xi, k_\perp) \biggr\vert_{3G} 
   \nonumber \\
   &&\quad= \frac{4\pi g_s^2 \xi^2} { (k_\perp^2)^2} 
        \int \frac{ d z}{z^2} d_{\bar q} (z) \frac{ 1}{z y^5} \biggr [ 2 (\xi^2 + 2x^2 z^2 -\xi xz) \biggr ( N(y,y) - O(y,y) \biggr ) 
\nonumber\\        
        &&\quad\quad   - 2 (\xi^2 + 4x^2 z^2 -3 \xi xz) \biggr ( N(y,0 ) + O (y,0) \biggr ) + y (\xi -x z)^2 \frac{ d}{dy} \biggr ( N (y,0)+ O (y,0) \biggr )
 \nonumber\\
       &&\quad\quad    - y (\xi^2 + x^2 z^2)  \frac{ d}{dy} \biggr ( N (y,y)   - O (y,y)  \biggr ) \biggr ], 
\label{FQT3G}                               
\end{eqnarray}     
with $y= x+ \xi/z$. 

The final result for $F_{qT}$ is the sum:
\begin{equation} 
  \frac{1}{m_A} F_{qT} (x,\xi,k_\perp) = \frac{1}{m_A}\biggr (    F_{qT}  \biggr\vert_{HP} +F_{qT}  \biggr\vert_{SGP} +  F_{qT} \biggr\vert_{SFP} + F_{qT}  \biggr\vert_{3G} \biggr )(x,\xi, k_\perp) ,   
\end{equation} 
where terms in the sum can be found in Eqs.(\ref{FQTHDP},\ref{FQTSGP}, \ref{FQTSFP},\ref{FQT3G}).

\par\vskip20pt
\noindent 
{\bf 4. Matching of $\Delta F_{qT}$} 

$\Delta F_{qT}$ represents the double spin asymmetry. Unlike $F_{qT}$, the asymmetry is not zero 
in the absence of absorptive parts in the scattering amplitude. 
In the matching of $\Delta F_{qT}$, there are contributions from Fig.\ref{TreeFq}. We first discuss the contribution from Fig.\ref{TreeFq}a.  It involves only gluon fragmentation function and can be written in the form: 
\begin{eqnarray}
{\rm Fig}.\ref{TreeFq}a   =  \int d^4 p   \frac{dz}{z^2}  \delta ( xP^+ - p^+ + k_g^+)  d_g (z) H_{kl} (p)   \int \frac{ d^4\xi}{(2\pi)^4} e^{-i \xi\cdot p} \langle h_A \vert  \bar q_k (\xi ) q_l (0) \vert h_A  \rangle,        
\end{eqnarray}    
where $k_g$ is the momentum of the gluon which is fixed as $z k_g =k$. $p$ is the momentum of the right quark line.  $H_{kl} (p)$ is the sum of the perturbative parts represented by the upper parts of the diagrams in Fig.\ref{TreeFq}a. With the power counting in Eq.(\ref{PWC})  for parton momenta we can expand $H_{kl} (p)$ around the momentum $\hat p^\mu = (p^+,0,0,0)$:
\begin{equation} 
 H_{kl} (p) =  H_{kl} (\hat p) + p_\perp^\alpha \frac{\partial H_{kl}}{\partial p_\perp^\alpha} ( \hat p) + \cdots,
 \label{HEXP}  
\end{equation}  
where $\cdots$ stand for contributions at higher orders. Taking the leading term and the twist-2 part of the quark density matrix in Eq. (\ref{TW2PDF}), one obtains the twist-2 contribution to $F_q$ and $\Delta F_q$
given as the first terms in Eq. (\ref{FQDFQ}).

\begin{figure}[hbt]
	\begin{center}
		\includegraphics[width=16cm]{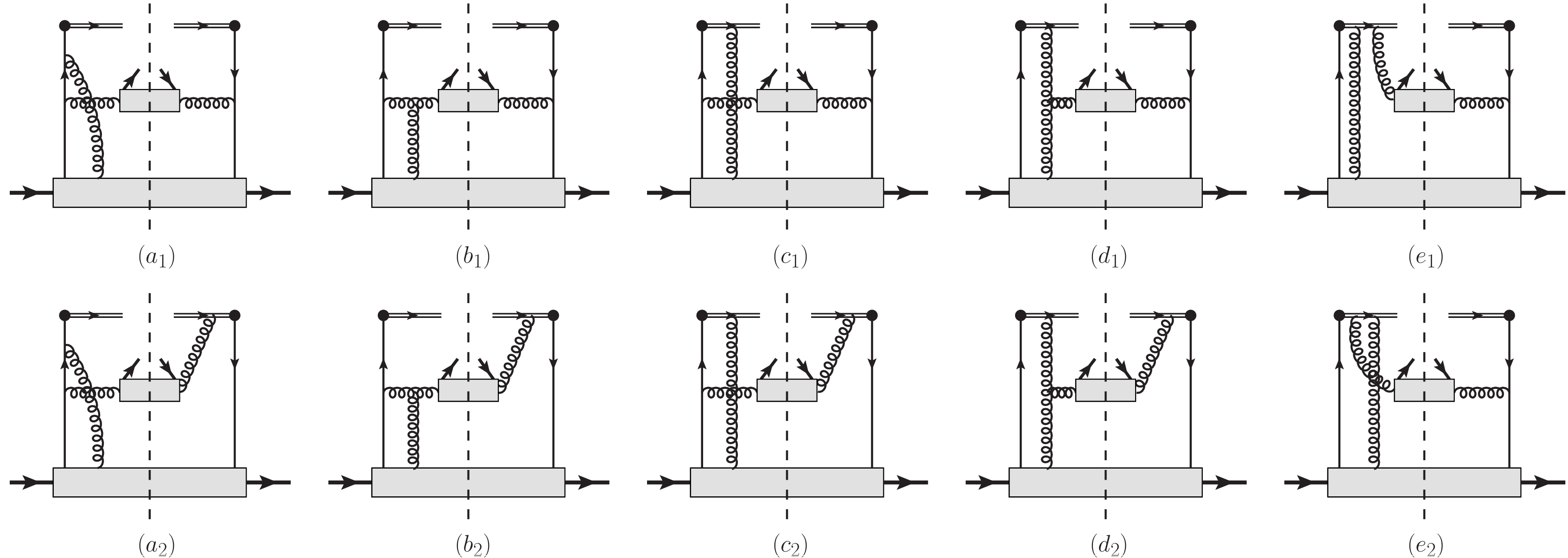}
	\end{center}
	\caption{Diagrams for the matching of $\Delta F_{qT}$. }
	\label{DFqt}
\end{figure}

The twist-3 contribution is obtained by taking the second term in the expansion in Eq.(\ref{HEXP}) or the first term combined with the twist-3 part of the quark density matrix. The result is:
\begin{eqnarray} 
&&\frac{2 k_\perp\cdot s_\perp  }{m_A} \Delta F_{qT}(x,\xi,k_\perp) \biggr\vert_{2a} \nonumber \\ 
&&\quad= 
   \frac{k_{\perp\mu}}{ (k_\perp^2 )^2 }  \int \frac{ d z}{z^2} d_g (z) \biggr [ 
 H_{2p, T} (x,\xi) \int \frac {d\lambda}{4\pi} P^+ e^{-i y P^+}\langle h_A  \vert \bar q (\lambda n) \gamma_\perp^\mu \gamma_5 q (0) \vert h_A \rangle 
\nonumber\\
    &&\quad\quad + H_{2p,\partial} (x,\xi)  \int \frac { d\lambda}{4\pi} i  e^{-i y P^+}\langle h_A \vert \bar q (\lambda n) \gamma^+  \gamma_5  \partial_\perp^\mu q (0) \vert h_A \rangle \biggr ] , 
\label{A2P} 
\end{eqnarray} 
with      
\begin{equation} 
    H_{2p, \partial} (x,\xi) = - 8 g_s^2 C_F \frac{\xi z^2  ( y^2 + 2 x^2)}{ y^3} , 
 \quad  H_{2p, T} (x,\xi) =  8 g_s^2 C_F \frac{\xi x^2 z^2 }{ y^2} , \quad y=x+\xi/z . 
\end{equation}         
These contributions seems to be with the twist-3 distributions $q_T$ and $q_\partial$, respectively.  But, they are not exactly those distributions. The contributions of gauge links are not included. When we consider the contributions in Fig.\ref{DFqt} where an additional gluon exchanges, 
parts of them will be the contributions of gauge links at the considered order.

In calculating the contributions with one-gluon exchange given by Fig.\ref{DFqt}, we have the same quark-gluon correlator given in Eq.(\ref{QGC}) and its  decomposition in Eq.(\ref{A2}).  
By calculating the contribution with $M_{A}^{\rho}$, we find that a part of the contribution can be added to the 
second term in Eq.(\ref{A2P}) so that the sum is obtained by inserting gauge links in the matrix element
at the considered order of $g_s$. 
With the gauge links  the matrix element is that used to defined $q_\partial$. Excluding this part, the contribution from Fig.\ref{DFqt} with $M_{A}^{\rho}$ is:
\begin{eqnarray} 
 &&\frac{2 k_\perp\cdot s_\perp }{m_A}  \Delta F_{qT} (x,\xi, k_\perp) \biggr\vert _{M_A^\rho} 
\nonumber\\
   &&\quad=   (-i g_s)  \int  d k_A^+  d k^+_1  \frac{ d z}{z^2} d_g (z)   \int \frac{d\lambda_1 d\lambda_2}{2 (2\pi)^2} e^{ -i\lambda_1 k_A^+ +i \lambda_2 k^+_1} \biggr \{   \frac{i}{ k^+_1}  H_{\perp A L \alpha}  (\hat k_A,\hat k_1)
\nonumber\\
   &&\quad\quad
    \langle h_A \vert \bar q ( \lambda_1 n    ) \gamma^+ \gamma_5  \biggr [\partial^+ G^{\alpha}_\perp -  \partial_\perp^\alpha G^+ 
     \biggr ](\lambda_2 n ) q( 0 )  \vert h_A   \rangle  + h.c. 
 \nonumber\\
     &&\quad\quad
 +  F_0 (x,\xi)  \frac{1}{ k^+_1}  k_{\perp} ^\alpha 
  \biggr [  \delta (k_A^+ - k_0^+)     \langle h_A  \vert   \bar q ( \lambda_1 n   ) \gamma^+ \gamma_5   G^+  (\lambda_2 n)   \partial_{\perp\alpha  }q(0  )  \vert h_A  \rangle
\nonumber\\
      &&\quad\quad
  + \delta (k_A^+ -k^+_1 -  k_0^+) 
    \langle h_A  \vert   \partial_{\perp\alpha  } \bar q ( \lambda_1 n   ) \gamma^+\gamma_5   G^+  (\lambda_2 n)  q( 0 )  \vert h_A  \rangle  \biggr  ] \biggr \} , 
\label{MALR}     
\end{eqnarray} 
with
\begin{equation} 
  \tilde H_{0}^\alpha (x,\xi) =  \tilde k_\perp^\alpha F_0 (x,\xi), \quad  H_{0}^\alpha (x,\xi) =  k_\perp^\alpha F_0 (x,\xi), \quad F_0 (x,\xi) = g_s^2 C_A \frac{\xi z^3 (\xi+ 2 xz)}{(k_\perp^2)^2 (\xi+ x z)^2 } . 
\end{equation}    
In Eq.(\ref{MALR}), the first two terms are gauge invariant, while the third- and fourth terms are not. The contribution from $M^\rho$ in the quark-gluon density matrix can be written:
\begin{eqnarray} 
&&\frac{2 k_\perp\cdot s_\perp }{m_A}  \Delta F_{qT} (x,\xi, k_\perp) \biggr\vert _{M^\rho} 
\nonumber\\
&&\quad =   g_s  \int  d k_A^+  d k^+_1  \frac{ d z}{z^2} d_g (z)   \int \frac{d\lambda_1 d\lambda_2}{2 (2\pi)^2} e^{ -i\lambda_1 k_A^+ +i \lambda_2 k^+_1} \biggr \{   \frac{i}{ k^+_1}  H_{\perp  L \alpha}  (\hat k_A,\hat k_1)
\nonumber\\
      &&\quad
    \langle h_A  \vert \bar q ( \lambda_1 n    ) \gamma^+   \biggr [ \partial^+ G^{\alpha}_\perp - \partial_\perp^\alpha G^+  \biggr ](\lambda_2 n ) q( 0 )  \vert h_A  \rangle  + h.c. 
 \nonumber\\
       &&\quad\quad
 +  F_0 (x,\xi)  \frac{1}{ k^+_1}  \tilde k_\perp^\alpha 
  \biggr [  \delta (k_A^+ - k_0^+)     \langle h_A  \vert   \bar q ( \lambda_1 n   ) \gamma^+   G^+  (\lambda_2 n)   \partial_{\perp\alpha  }q(0  )  \vert h_A \rangle
\nonumber\\
       &&\quad\quad
  + \delta (k_A^+ -k^+_1 -  k_0^+) 
    \langle h_A \vert   \partial_{\perp\alpha  } \bar q ( \lambda_1 n   ) \gamma^+   G^+  (\lambda_2 n)  q( 0 )  \vert h_A  \rangle  \biggr  ] \biggr \} , 
\label{MLR}     
\end{eqnarray} 
where the first two terms are gauge invariant. The third- and fourth terms are not gauge invariant. It is interesting to note that they have the same perturbative coefficient function $F_0$ as that in the gauge variant contribution in Eq.(\ref{MALR}).  

Now we consider the contributions with $M_\perp^{\rho\mu}$ and $M_{A\perp}^{\rho\mu}$ in the quark-gluon density matrix. The twist-3 contributions are given by taking the index $\rho=+$, i.e, only the field component $G^+$ is involved.  We find that the contribution from $M_\perp^{\rho\mu}$ is proportional to $C_A$. 
In the contribution from $M_{A\perp}^{\rho\mu}$, there is a part proportional to $C_F$. This part can be summed with the first term in Eq.(\ref{A2P}). The sum is obtained by inserting gauge links in the matrix element in the first term so that the matrix element is that used to defined $q_T$. The sum is then gauge invariant at the considered order of $g_s$. 
The sum of the  remaining contribution from $M_{A\perp}^{\rho\mu}$ and that from $M_{\perp}^{\rho\mu}$ can be written in the form: 
 \begin{eqnarray} 
\frac{2 k_\perp\cdot s_\perp }{m_A}  \Delta F_{qT} (x,\xi, k_\perp) \biggr\vert _{C_A} 
 &=&    g_s \int \frac{ dz}{z^2} d_g (z) F_0 (x,\xi)  \int d k_A^+ d k^+_1  \biggr ( - 1 + \frac{k_0^+}{ k^+_1} \biggr )  \delta (k^+_A - k_0^+) 
\nonumber\\
    &&  \int \frac{d\lambda_1 d\lambda_2}{2 (2\pi)^2} e^{-  i\lambda_1 k_A^+ +i \lambda_2 k^+_1}  
     \langle h_A  \vert   \bar q ( \lambda_1 n   )   \biggr ( \gamma\cdot k_\perp \gamma_5 
 \nonumber\\    
    &&  + i \gamma \cdot \tilde k_\perp \biggr )    G^+  (\lambda_2 n)  q( 0 )  \vert h_A )  \rangle  + h.c. .
\label{CA}       
\end{eqnarray}       
This sum is not gauge invariant and it involves the same perturbative function $F_0$. 
We notice that the quark field can be written as the sum of a $+$- and $-$-component:
\begin{equation} 
  q^{(+)}(x) = \frac{1}{2} \gamma^- \gamma^+ q(x), \quad q^{(-)}(x) =\frac{1}{2} \gamma^+ \gamma^- q (x), \quad q (x) = q^{(+)} (x) + q^{(-)}(x). 
 \end{equation}  
The $-$-component is not independent. With the equation of motion one has: 
\begin{equation} 
   q^{(-)}(x) = \frac{1}{2} {\mathcal L}_n^\dagger (x)  \int_0^{\infty} d \lambda \biggr ( {\mathcal L}_n \gamma^+\gamma_\perp^\mu  D_\mu  q^{(+)} \biggr ) ( \lambda n +x) . 
 \label{EOMS}   
\end{equation}    
In Eq.(\ref{CA}) one quark field has to be the $-$-component. Using the solution in Eq.(\ref{EOMS}) 
one has the sum: 
 \begin{eqnarray} 
\frac{2 k_\perp\cdot s_\perp }{m_A}  \Delta F_{qT} (x,\xi, k_\perp) \biggr\vert _{C_A}    &=& g_s \int \frac{ dz}{z^2} d_g (z) F_0 (x,\xi)  \int d k_A^+ d k^+_1   \frac{i}{k^+_1}   \int \frac{d\lambda_1 d\lambda_2}{2 (2\pi)^2} e^{-  i\lambda_1 k_A^+ +i \lambda_2 k^+_1}       
\nonumber\\
    &&  \biggr  [ \delta (k^+_A - k_0^+)      \langle h_A  \vert   \bar q ( \lambda_1 n   )   \biggr (  k_\perp^\alpha \gamma^+  \gamma_5 
  + i \tilde k_\perp^\alpha \gamma^+  \biggr )    G^+  (\lambda_2 n)  \partial_{\perp\alpha} q( 0 )  \vert h_A  \rangle 
\nonumber\\
    &&  + \delta (k_A^+ - k^+_1 -k_0^+ )       \langle h_A  \vert  \partial_{\perp\alpha}  \bar q ( \lambda_1 n    )   \biggr (  k_\perp^\alpha \gamma^+  \gamma_5 
\nonumber\\    
   &&  -  i \tilde k_\perp^\alpha \gamma^+  \biggr )    G^+  (\lambda_2 n)   q( 0 )  \vert h_A  \rangle \biggr ].  
\label{LSP}        
\end{eqnarray}    
Comparing the sum with the gauge variant contribution in Eq.(\ref{MALR}, \ref{MLR}), we find that all gauge 
variant contributions are cancelled each other so that the remaining contribution is gauge invariant. We have then the result which is gauge invariant: 
\begin{eqnarray} 
\frac{1}{ m_A} \Delta F_{qT} (x,\xi, k_\perp)\biggr\vert_{q\bar q + qG\bar q}  &=&\frac{1}{2  (k_\perp^2)^2} \int \frac{ dz}{z^2} d_g(z)  \biggr [ \biggr ( H_{2p, T} (x,\xi) q_T (y) 
  + H_{2p,\partial} (x,\xi) q_\partial (y) \biggr )
\nonumber\\
    && + \frac{2}{\pi}  \int  d x_2  \biggr ( T_F (y, x_2) H (x,\xi, x_2) + T_\Delta (y,x_2) H_A(x,\xi, x_2 ) \biggr ) \biggr ] ,  
\label{E61}          
\end{eqnarray}  
with $H_{2p,\partial}$ and $H_{2p, T} $ are given in Eq.(\ref{A2P}).  $H(x,\xi, x_2 )$ and $H_A(x,\xi,x_2) $ 
are given as: 
\begin{eqnarray} 
H(x,\xi,x_2) &=& \frac{2 g_s^2\xi z^2} {y x_2 (y-x_2)} \biggr [ C_A\frac{x^2-x_2 y}{x_2-x} + 2 C_F \frac{1}{y} 
   (xy -x_2(x+y)) \biggr ], 
\nonumber\\
H_A (x,\xi,x_2) &=& \frac{2 g_s^2 \xi z^2} { y x_2 (x_2-y)} 
  \biggr [ C_A \frac{(x^2 +x_2 y)(x_2+y-2x) }{(x_2-x)(x_2-y)}  + 2 C_F \frac{1}{y} (x (2x-y) + x_2 (x+y))\biggr].    
\end{eqnarray}  

\par
$\Delta F_{qT}$  receives contributions from twist-3 gluon distributions, where the antiquark fragmentation function is involved. These contributions are given by diagrams of two-gluon exchanges as given in Fig.\ref{TreeFq}b and by three-gluon exchanges in Fig.\ref{G3} without the cut on the quark propagator and the gauge link. It is noted that the contribution of three-gluon exchanges from the second- and third diagram in Fig.\ref{G3}a is the same. Because of Bose-symmetry, one should either take any one of the two diagrams,  or the half of the sum into account.

The contribution of two-gluon exchanges from Fig.\ref{TreeFq}b can be written in the form: 
\begin{eqnarray} 
{\rm Fig.\ref{TreeFq}b} 
    =  \int \frac{d z}{z^2} d_{\bar q}  (z)  d^4 k_1 H^{\mu\nu} (k_1) \int \frac{d^4 x}{(2\pi)^4} 
         e^{ix\cdot k_1} \langle h_A  \vert  G^{a}_\mu(0) G^{a}_\nu (x)  \vert h_A \rangle, 
\end{eqnarray}          
where $H^{\mu\nu} (k_1)$ is the sum of the perturbative parts represented by the upper parts of the diagrams in Fig.\ref{TreeFq}b. $k_1$ is the momentum carried by the gluons. 
It is easy to check the following Ward identities:
\begin{equation} 
   k_{1\mu} H^{\mu\nu} (k_1) =0, \quad k_{1\nu}  H^{\mu\nu} (k_1) =0. 
\label{WI2}    
\end{equation}    
Before doing the collinear expansion in $k_1$ around $\hat k_1^\mu = (k_1^+, 0,0,0)$, we can use these identities 
to manipulate the expression:
\begin{eqnarray} 
 && H^{\mu\nu} (k_1) \int \frac{d^4 x}{(2\pi)^4} 
         e^{ix\cdot k_1} \langle h_A \vert  G^{a}_\mu(0) G^{a}_\nu (x)  \vert h_A\rangle
\nonumber\\
  && = \frac{1}{ (k_1^+)^2}\int \frac{d^4 x}{(2\pi)^4}  e^{ix\cdot k_1}   \biggr ( H^{\mu_\perp \alpha_\perp} (k_1) \langle h_A \vert   \hat G^{a,+}_{\quad \ \mu_\perp} (0)  \hat G^{a,+}_{\quad\  \alpha_\perp}  (x)  \vert h_A\rangle 
\nonumber\\  
   && \quad  +  H^{+ \alpha_\perp} (k_1) \langle h_A \vert  \hat G^{a,+ - } (0)  \hat G^{a,+}_{\quad\  \alpha_\perp}  (x)  \vert h_A  \rangle  +H^{\mu_\perp  +} (k)   \langle h \vert   \hat G^{a,+ }_{\quad \ \mu_\perp } (0)  \hat G^{a,+-} (x)  \vert h_A \rangle 
 \nonumber\\
     && \quad   + H^{+ +} (k_1)   \langle h_A \vert   \hat G^{a,+- }(0)  \hat G^{a,+-} (x)  \vert h_A \rangle \biggr ) , 
\end{eqnarray}         
where $ \hat G^{\mu\nu} = \partial^\mu G^\nu - \partial^\nu G^\mu$.                   
In the first term, the leading contribution in the collinear expansion of $H^{\mu_\perp \alpha_\perp} (k_1)$ gives the twist-2 contribution, i.e., to $\Delta F_q$.   The twist-3 contribution is obtained by expanding the next-to-leading contribution. The leading contribution of the second- and third term is at twist-3. The last term is a twist-4 contribution which can be neglected. In the collinear expansion of $H^{\mu\nu}$, we find that the next-to-leading contribution  $H^{\mu_\perp \alpha_\perp} (k_1)$ is antisymmetric in the indices $\mu_\perp$ and $\alpha_\perp$. Because of symmetries, the related matrix element is symmetric in the indices $\mu_\perp$ and $\alpha_\perp$, as shown in Eq.(\ref{MPGG}). Hence, the first term gives no contribution. 
Therefore, only the second- and third gives nonzero contribution at twist-3. The contribution can be expressed by the twist-3 gluon distribution: 
\begin{equation} 
 g_{3T} (x) \tilde s^\mu =   \frac{i}{x} \int \frac{ d\lambda}{2\pi}
e^{ - i x \lambda  P^+ }
    \langle h_A   \vert   G^{+ -  } (\lambda n  ) {\mathcal L}_n^\dagger  (\lambda n) 
      {\mathcal L}_n (0) G^{+\mu }(0)  \vert h_A  \rangle. 
\label{GHAT}      
\end{equation} 
We obtain the contribution from Fig.\ref{TreeFq}b:
\begin{equation} 
\frac{1}{m_A} \Delta F_{qT} (x,\xi, k_\perp)\biggr\vert_{2G} = \frac{16\pi \alpha_s }{(k_\perp^2)^2} \int \frac{ dz}{z^2} d_{\bar q} (z) \frac{ \xi^2 x z}{y^2} g_{3T} (y), \quad  y = x + \frac{\xi}{z}. 
\label{TWOG} 
\end{equation} 
It is noted that from Fig.\ref{TreeFq}b we only obtain the contribution with $g_{3T}$ defined only 
with $\hat G^{\mu\nu}$ and without gauge links. The contribution alone is  not gauge invariant.   
There is a difference of a two-gluon term 
between $\hat G^{\mu\nu}$ and $G^{\mu\nu}$.   
In the calculation of three-gluon exchanges given by diagrams in Fig.\ref{G3}, we find that a part of the contribution from Fig.\ref{G3} gives the needed two-gluon term in $G^{\mu\nu}$ and another part 
forms the gauge links in $g_{3T}$ at the considered order of $g_s$.

The contribution from three-gluon exchanges is given by diagrams in Fig.\ref{G3}. It can be written 
\begin{eqnarray} 
{\rm Fig}.\ref{G3} &=&  \int \frac{ dz}{z^2} d_{\bar q} (z)  \int d^4 k_1 d^4 k_2 d^4 k^4_3 H_{\mu_1\mu_2\mu_3}^{abc}  (k_1,k_2,k_3) 
\nonumber\\
  &&  \int \frac{d^4 x_1 d^4 x_2 d^4 x_3} {(2\pi)^{12}}  e^{i k_1 \cdot x_1 + ik_2\cdot x_2 - i k_3 \cdot x_3} 
   \langle h_A \vert G^{c,\mu_3} ( x_3) G^{b,\mu_2} (x_2) G^{a,\mu_1} (x_1) \vert h_A\rangle, 
\end{eqnarray} 
where $H_{\mu_1\mu_2\mu_3}^{abc}(k_1,k_2,k_3) $ is the sum of the perturbative parts represented by the upper parts of the diagrams in Fig.\ref{G3}.  The two gluons in Fig.\ref{G3}a carry the momentum $k_1$ and $k_2$, respectively. $k_3$ is the momentum carried by the gluon in Fig.\ref{G3}b. To obtain the gauge invariant result, we first note that there is a Ward identity related to the gluon in Fig.\ref{G3}b:
\begin{equation} 
    k_3^{\mu_3}H_{\mu_1\mu_2\mu_3}^{abc}(k_1,k_2,k_3)  =0. 
\end{equation}       
This identity can be easily checked as those in Eq.(\ref{WI2}). With this identity we can write the contribution 
as:
\begin{eqnarray} 
{\rm Fig}.\ref{G3} &=&  \int \frac{ dz}{z^2} d_{\bar q} (z)  \int d^4 k_1 d^4 k_2 \frac{i}{k_3^+}  H_{\mu_1\mu_2\mu_3}^{abc}  (k_1,k_2,k_3) 
\nonumber\\
  &&  \int \frac{d^4 x_1 d^4 x_2 } {(2\pi)^{8}}  e^{i k_1 \cdot x_1 + ik_2\cdot x_2 } 
   \langle h_A \vert \hat G^{c,+\mu_3} ( x_3) G^{b,\mu_2} (x_2) G^{a,\mu_1} (x_1) \vert h_A \rangle, 
\end{eqnarray} 
with $k_3 = - k_1 -k_2$. To proceed further, we notice that the perturbative coefficient function $H_{\mu_1\mu_2\mu_3}^{abc}$ can be decomposed into a totally symmetric and a totally asymmetric part in the three color indices: 
\begin{equation} 
H_{\mu_1\mu_2\mu_3}^{abc}  (k_1,k_2,k_3) = H_{\mu_1\mu_2\mu_3}^{abc}  (k_1,k_2,k_3)\biggr\vert_{d} + H_{\mu_1\mu_2\mu_3}^{abc}  (k_1,k_2,k_3)\biggr\vert_{f} , 
\label{HSA} 
\end{equation} 
where the first term is the symmetric part proportional to $d^{abc}$, and the second term is antisymmetric part proportional to $f^{abc}$. For the symmetric part, we have still the following Ward identities 
\begin{equation} 
 k_1^{\mu_1} H_{\mu_1\mu_2\mu_3}^{abc}  (k_1,k_2,k_3)\biggr\vert_{d}=0 , \quad 
k_2^{\mu_2} H_{\mu_1\mu_2\mu_3}^{abc}  (k_1,k_2,k_3)\biggr\vert_{d} =0 , 
\label{SWI} 
\end{equation} 
because of that the color symmetric part is essentially an amplitude of QED. With these identities one can 
write the contribution from the symmetric part in a gauge invariant form. 

\par 
  
For the antisymmetric part there are no Ward identities like those in Eq.(\ref{SWI}).  We need to do the collinear expansion first and then try to find a gauge invariant result. The calculation of this part is very tedious. One has to expand the antisymmetric part  in Eq.(\ref{HSA}) to order of $\lambda^2$ in order to obtain the complete twist-3 contribution. At the end, we find that a part of the antisymmetric part gives the contributions which can be combined with the contribution of two-gluon exchange into a gauge invariant form as discussed after Eq.(\ref{TWOG}). The remaining contribution of the antisymmetric part can be written in the form with the three-gluon correlation function $M_F^{\mu\nu\rho}$ defined in Eq.(\ref{TW3GM}).  
We have the following result from Fig.\ref{G3}:    
\begin{eqnarray} 
\frac{1}{m_A} \Delta F_{qT} (x,\xi,k_\perp) \biggr\vert_{3G} &=& \frac{g_s^2}{(k_\perp^2)^2} 
\int \frac{ dz}{z^2} d_{\bar q} (z) \int d x_2  \frac{ -4 \xi^2}{x_2^2 (y-x_2) y^3}  \biggr \{ x_2 \xi \biggr [ O (y-x_2,y)
\nonumber\\ 
  &&  - N(y-x_2,y) \biggr ]   + ( 2 \xi y + y x_2 z - y^2 z -\xi x_2) \biggr [ N (y_2, x_2) + O(y, x_2) \biggr ] 
\nonumber\\
   && +  y( z(y+ x_2) -2 \xi ) \biggr [ N (y-x_2,-x_2) + O(y-x_2,-x_2)\biggr ] \biggr \} . 
\label{THRG}    
\end{eqnarray}     
The total matching result is the sum of the results  in Eq.(\ref{TWOG},\ref{THRG}). It is noted that the twist-3 
distribution $g_{3T}$ in Eq.(\ref{TWOG}) is not independent. Using equation of motion and the relation in Eqs.(\ref{RL3G}, 26),   $g_{3T}$ can be expressed with  
the twist-3 distribution $T_F$ and $N$. We have:
\begin{equation} 
 g_{3T} (x) = -\frac{1}{x^2} \int d x_2\biggr \{\frac{1}{\pi} T_F (x_2 +x, x) + P\left (\frac{1}{x-x_2} \right ) 
 \biggr [ N(x,x-x_2) - N(x,x_2) + 2 N(x_2,x_2-y) \biggr ] \biggr \}. 
\end{equation}    
The result of $\Delta F_{qT}$ from Fig.\ref{TreeFq}b and Fig.\ref{G3} is:
\begin{eqnarray} 
&&\frac{1}{m_A} \Delta F_{qT} (x,\xi, k_\perp) \biggr\vert_{2G+3G }  
\nonumber\\ 
  &&\quad= \frac{-16\pi \alpha_s \xi^2 }{(k_\perp^2)^2 } \int \frac{ dz d x_2}{z^2 y^3} d_{\bar q} (z) \biggr \{ \frac{xz}{\pi y} T_F (x_2,x_2+y) -\frac{2 xz}{y(y-x_2) } 
\biggr [ N(y,x_2) 
\nonumber\\
 &&\quad\quad - N(y-x_2,y) + 2 N(y-x_2,-x_2)  \biggr]  +\frac{1}{x_2^2 (y-x_2) } \biggr ( x_2 \xi \biggr [ O(y-x_2,y) 
\nonumber\\   
     &&\quad\quad  - N(y-x_2,y) \biggr ] 
     + ( 2 \xi y + y x_2 z -y^2 z -\xi x_2) \biggr [ N(y,x_2) + O(y,x_2) \biggr ] 
\nonumber\\
          &&\quad\quad  + y( z(y+x_2) -2 \xi) \biggr [ N(y-x_2,-x_2) + O (y-x_2,-x_2) \biggr ] \biggr ) \biggr \} . 
\label{E75}          
\end{eqnarray}  
The complete matching result is the sum of that in Eq.(\ref{E61}) and  that in Eq.(\ref{E75}):         
\begin{equation} 
\frac{1}{ m_A} \Delta F_{qT} (x,\xi, k_\perp) = 
\frac{1}{ m_A} \Delta F_{qT} (x,\xi, k_\perp)\biggr\vert_{q\bar q + qG\bar q} +
\frac{1}{ m_A} \Delta F_{qT} (x,\xi, k_\perp)\biggr\vert_{2G+3G} . 
\end{equation} 
   
\par\vskip20pt

\noindent 
{\bf 5. Summary} 
\par 

SIDIS in target fragmentation region can be conveniently described with fracture functions, i.e., one can use QCD factorization with fracture functions to make predictions in this region.  If the transverse momentum of the produced hadron is in the region $Q\gg k_\perp \gg\Lambda_{QCD}$,  the standard collinear factorization can also be used. Therefore, 
fracture functions can be factorized or matched with parton distributions and fragmentation functions. We have studied the matching up to twist-3 level.   At the order of $\alpha_s$ considered in this work, fracture functions are only matched to twist-2 parton fragmentation functions with twist-2- or twist-3 parton distributions.  There is no contribution from chirality-odd parton distributions. We have derived perturbative coefficient functions in the factorization or matching of twist-2- and twist-3 fracture functions. Especially in the derivation of twist-3 fracture functions, we find that the results can be written in a gauge invariant form. 
These results will be useful for modeling of fracture functions and resummation of large logarithms of $k_\perp$ in collinear factorization.

\par\vskip40pt
% The correct funding number:  12075299 (mian shang 2021-2024) 
% qunti 11821505, penghuanwu 12047503
% Adding lattice funding number 11935017(2020-2024)
% The number 11675241 is for 1017-2020. 

\noindent
{\bf Acknowledgments}
\par
The work is supported by National Natural Science Foundation of China(No.12075299,11821505, 11935017 and 12047503)  and by the Strategic Priority Research Program of Chinese Academy of Sciences, Grant No. XDB34000000. K.B. Chen is supported by National Natural Science Foundation of China (Nos. 11947055, 12005122) and Shandong Province Natural Science Foundation No. ZR2020QA082. X. B. Tong is supported by the CUHK-Shenzhen university development fund under Grant No. UDF01001859.

\par\vskip40pt
\par

\end{document}